\documentclass[mathleft]{an}
\usepackage{times}
\usepackage{graphicx}

\begin{document}
%
% The following seven commands are intended for editorial usage and should be ignored by
% the author(s).
\Pagespan{789}{}% Document's page range. 
% If second parameter is left empty, the last page is computed automatically.
\Yearpublication{2007}%
\Yearsubmission{2007}%
\Month{11}%   
\Volume{999}%  
\Issue{88}% 
% \DOI{This.is/not.aDOI}% 

   \title{Results and Perspectives of Young Stellar Object long look programs}
   
   \author{S. Sciortino \inst{1}\fnmsep\thanks{Corresponding author:
 \email{sciorti@astropa.inaf.it}\newline}}

   \institute{INAF-Osservatorio Astronomico di Palermo Giuseppe S. Vaiana,
Piazza del Parlamento 1, 90134 Palermo, Italy \\ 
             }
\received{3 Sept 2007}
\accepted{?? ?? 2007}
\publonline{later}


\abstract{Both {\em Chandra} and {\em XMM-Newton} have performed long look programs for studying the YSO physics. I will discuss recent results on the controversial issue of Class~0 YSO X-ray emission, the observational evidence of magnetic funnels interconnecting the YSO with its circumstellar disk and the Fe 6.4 keV fluorescent line emission and its origin.
While recent results of the XMM-Newton {\em DROXO} program challenge the "standard" interpretation of the Fe 6.4 line origin as due to photoionized fluorescing disk material, the discovery of X-ray excited Ne 12.81 $\mu$m line is a clear evidence of the interaction between X-rays and disk material. Future long look observations with XMM-Newton are required to clarify the X-ray effects on YSO disk.}
  
   \authorrunning{S. Sciortino}
   \titlerunning{Young Stellar Object long look programs}
   \maketitle
%
%________________________________________________________________

\section{Introduction}

X-ray emission likely traces, and is related to, magnetic
fields at work in the interaction region between the central Young Stellar Object (YSO) and its surrounding disk. Because of their role, X-rays have started to be recognized as an important element for understanding star 
formation and early evolution.
Since the launch of {\em Chandra} and {\em XMM-Newton} X-ray emission from YSOs has been the subject of many studies focussed on nearby Star Forming Regions (SRFs), among those are notable 
few selected long look programs. So far two of these programs have been performed with {\em Chandra}: i) {\em COUP}
(\protect{{\em C}handra} {\em O}rion {\em U}ltradeep {\em P}roject, PI: E. Feigelson,
cf. \cite{ss_Getman+05}), a 850 ks long continuous observation of the Orion Nebula Cluster region which has allowed us to study the X-ray properties of known Orion YSOs as well as to discover and characterize the Orion YSO embedded population, and ii) a 450 ks long observation of the young cluster N~1893 in the external side of the Galaxy. This latter program, led by G. Micela, aims to study the IMF in the external region of the Galaxy where the environmental conditions are different than in the vicinity of the Sun. I will not discuss anymore this latter program since I will concentrate on the role of X-rays on YSO physics and evolution. In this specific realm only one long look program, led by myself, has been performed with XMM-Newton, it is nicknamed {\em DROXO} ({\em D}eep {\em R}ho {\em O}phiuchi {\em
X}MM-Newton {\em O}bser-vation). It consists of a 500 ks long continuous observation of the $\rho$ Oph core F region (\cite{ss_Sciortino+06}). Thanks to XMM-Newton high throughput the {\em DROXO} time resolved spectroscopy is allowing us to study the X-ray emission of the 1 Myr old 
$\rho$ Oph YSOs and its impact on YSO physics.

Another extensive program devoted to the study of YSO physics is {\em XEST} ({\em X}MM-Newton {\em E}xtended {\em S}urvey of {\em T}aurus, \cite{ss_Gudel+07}) that has adopted an observational strategy different from a long look one. The program and its results are presented in this volume by M. G\"udel. 

In the current scenario of Class~I-II YSOs (Hartmann 1998) magnetically funneled accretion streams connect the central star with its circumstellar disk. In such a system X-rays could be emitted by the PMS star corona, by the funnel plasma that is shocked as it accretes on the star, by the fluorescing disk matter or by gas shocked in a jet. Some of the recent observations, with crucial contributions from {\em COUP} and {\em XEST}, have shown evidence that all the above contributions can indeed be present. However from an observational point of view, we need a stronger and more compelling evidence, i.e. to find a way to distinguish, recover and study the properties of those distinct contributions. On more general ground X-rays are likely to be crucial to understand the chemistry and evolution of proto-planetary disks. In the following I will briefly discuss some of the current open issues on YSO physics.

\begin{figure}
 \centering
% \resizebox{\hsize}{!}{
 \includegraphics[width=6.5cm,height=6.5cm]{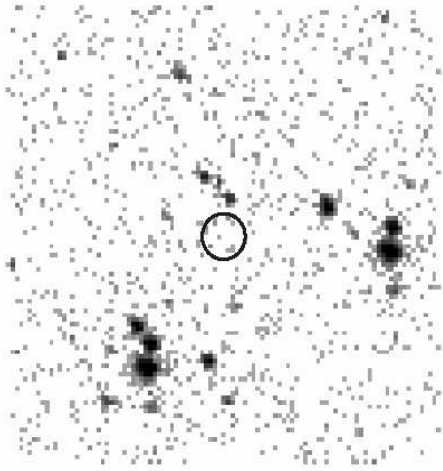}%}
 \caption{Stacked ACIS image from events collected in the energy interval
$\Delta$E = 0.5$-$8.0 keV taken from six regions of
200 $\times$ 200 pixels centered on the position of the 6 known Class~0 sources in the Serpens.
The circular region in the center is 5", corresponding to positional
uncertainties of the known mm/submm sources. 
No X-ray excess has been found within this area indicating that the
Serpens Class~0 YSOs are unlikely
to be X-ray sources with intensities just below the detection threshold
(adapted from \cite{ss_Gia+07a}).}
\label{fig6_ss}
\end{figure}

\section{X-rays from Class 0 YSOs}

\begin{figure}[h]
 \centering
 \includegraphics[width=6.5cm]{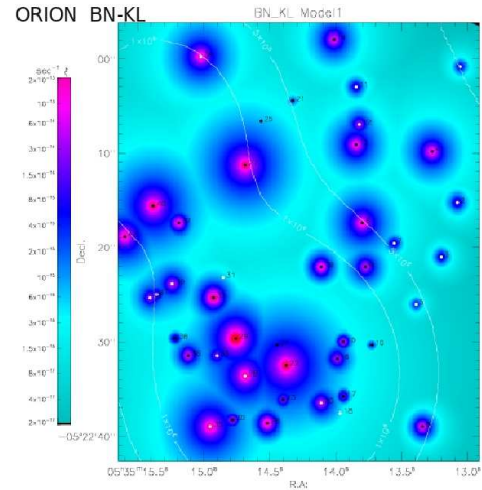}
 \caption{Two dimensional projection of a model computation of the (color coded)
ionization rate as function of position for the BN cloud core. Across the entire core the ionization rate is higher than predicted by cosmic 
rays (2 $cdot$ 10$^{-17}$ s$^{-1}$). Around each of the 
embedded X-ray emitting YSOs develops a R\"ontgen sphere where the 
X-ray induced
ionization rate is several orders of magnitude higher than the "background" level (courtesy of A. Lorenzani and F. Palla.).}
\label{fig6b_ss}
\end{figure}

After many years of search, the occurrence of X-ray emission from Class~0 YSOs is still controversial. Either this
emission is weak or rare or it is hidden due to
the conspicuous amount of intervening absorbing material.
In fact, while X-rays are quite
penetrating -- indeed the absorption at 2 keV and at 2 $\mu$m are
similar (\cite{ss_Reyter96})-- Class~0 sources can be subject to extinction up to hundreds of
magnitudes preventing the escape of any X-rays.  
One of the most (if not the most) stringent upper limit to the intrinsic 
X-ray luminosity of Class~0 has been obtained thanks
to a 100 ksec {\it Chandra} observation toward the
Serpens SFR (\cite{ss_Gia+07a}). By staking data taken at 6 known Class~0 
positions, i.e. by constructing a virtual $\sim$ 600 ks long observation, the  Class~0 intrinsic X-ray luminosity has resulted to be lower than
 4~10$^{29}$ erg/s (assuming
emission from an optically thin isothermal plasma with kT = 2.3 keV seen
through an absorbing column with N$_H$ = 4~10$^{23}$ cm$^{-2}$). However the best upper limit so far obtained is still a dex higher than the X-ray luminosity of active
Sun. Future deep observations are needed to really advance 
our knowledge on this subject that could affect our understanding of
star formation process. In fact with COUP we have discovered a deeply embedded population in Orion (\cite{ss_Grosso+05}), that has been
shown to locally dominate the ionization level within the given
molecular cloud core (cf. \cite{ss_Lorenzani+07}, and Fig. \ref{fig6b_ss}).
Still, as of today, we do not know when intense X-ray emission from
YSOs really develops likely affecting --for example by
determining the effectiveness of ambipolar diffusion -- the further evolution of star formation process. We do not know yet if this effect is
just a small adjustment of the current interpretational 
scenario(s) or a major change is required if X-rays start acting at very early (Class~0) times.

\section{Flares and Magnetic Funnels}

X-ray flares are a classic tool to derive physical parameters of an emitting region (cf. Reale 2007). In fact the use of dynamical information (decay time, etc.) allows deriving physical characteristics of the flaring region. This is possible because in order to have a flare with the typical decay phase the plasma 
{\bf must} be confined (\cite{ss_Reale+02}). As a results the behavior of flare light curve (and the related time resolved spectra) allows measuring the size of flaring magnetic structure. In normal stars the observed flares are similar to solar ones,
but sometimes much stronger (up to 10$^4$), both in absolute terms and with respect to the star bolometric luminosity. In most cases the observed YSO flares fall in the same category, but there are a few notable exceptions: in about 10 {\em COUP} (\cite{ss_Favata+05a}) and 2 {\em DROXO} 
(\cite{ss_Flacc+07a}) flares, the analysis results in a size of the flaring region a least 3 times larger than stellar radius and in few cases as long as 0.1 AU, i.e. the size of the star-disk separation. These long structures have never been seen in more evolved normal stars. Such long structures, if anchored on the stellar surface, will suffer severe stability problem
due to the centrifugal force -- 1-2 Myr YOSs are fast rotators (P$_{rot}$ = 1-8 days) with a disk corotation radius of about 1-10 stellar radius --  hence they would be ripped open. A possible alternative scenario is one in which the loop connects the star with the disk at the corotation radius. This is compatible with the currently available observational evidence. Such magnetic funnels have been predicted by magnetospheric accretion models
(e.g., \cite{ss_Shu+97}) and have been shown to occur in 
up-to-date MHD simulations of disk-star system (eg., \cite{ss_Long+07}), but it is only thanks to the {\em COUP} and {\em DROXO} long look observations that we have gained some observational evidences of their existence.

\begin{figure}
 \centering
 \includegraphics[width=6.5cm]{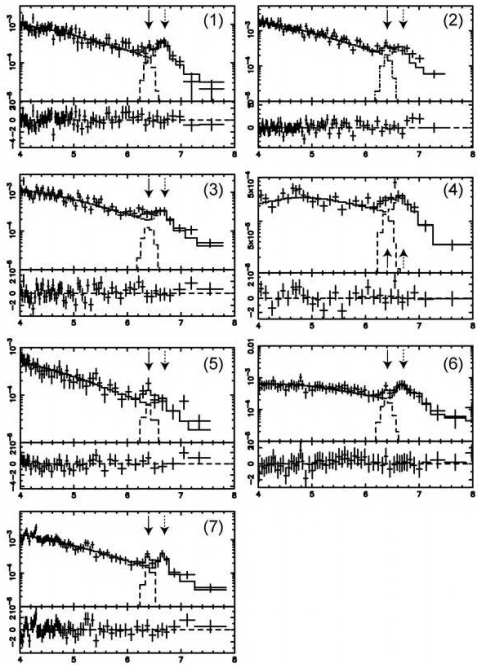}
 \caption{(Upper panels) Observed spectra (pluses) and best-fit models
(solid steps) of the 7 COUP YSOs showing the Fe fluorescent 6.4 keV
line. The Fe 6.4 keV line Gaussian component is shown by dashed steps.
The 6.4 and 6.7 keV line position are indicated by solid and broken arrows,
respectively. Photon energy in keV is on the abscissa, while
the ordinate is the spectral intensity as counts s$^{-1}$
keV$^{-1}$. (Lower panels) Residual to the fit in unit of $\chi$ values
(adapted from \cite{ss_tu+05}).}
\label{fig7_ss} 
\end{figure}

\section{The Fe 6.4 keV fluorescent line emission of YSOs}

The first detection of the Fe 6.4 keV fluorescent line in a YSO
has been obtained with Chandra during an intense flare on YLW16A, a Class~I YSO 
in the $\rho$ Oph SFR (Imanishi et al. 2001).
Thanks to COUP we have 
collected 134 Orion YSO good quality spectra that allows investigating the presence of the
$\sim$ 6.4 keV Fe K$_{\alpha}$ line.  In 7 COUP sources the 6.4 keV line
(cf. Fig. \ref{fig7_ss}) has been found (\cite{ss_tu+05}) and the emission
has been interpreted, following original suggestion of \cite{ss_im+01},
as due to the circumstellar neutral disk matter illuminated by
the X-rays emitted from the PMS star during the intense flares observed in all those seven sources. A Fe 6.4 keV fluorescent line has
also been seen during a relatively short XMM-Newton observation of the
Class~II YSO Elias 29 without any evidence of concurrent flare emission (\cite{ss_Favata+05b}).
In all the above reports none or very limited time resolved spectroscopy
has been possible due either to the XMM-Newton too short observation or
the {\it Chandra} limited collecting area. Very recently \cite{ss_Czesla+Schmitt07}
have reported the results of time-resolved spectroscopy of V1489 Ori, one of the 7 COUP 
sources with the Fe 6.4 eV line, showing that the K$_{\alpha}$ line appears 
predominantly during the 20 ks rise phase of a flare. Their initial calculation 
suggests that the photo-ionization alone cannot account for the observed
intensity of the Fe K$_{\alpha}$ line

\begin{figure*}
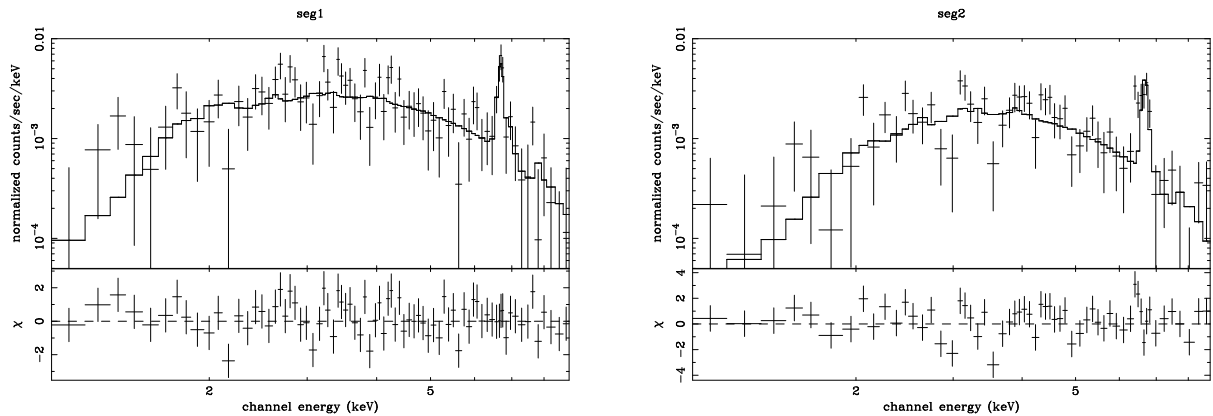

 \centering
\resizebox{16cm}{!}
{{\rotatebox[]{-90}{\includegraphics[width=6cm,clip=true]{SSc_El29_seg1.ps}}}
\hspace{1cm}
{\rotatebox[]{-90}{\includegraphics[width=6cm,clip=true]{SSc_El29_seg2.ps}}}}
\vspace{-1.1cm}
 \caption{Spectra and spectral fit to the {\em DROXO} data of Elias 29 before the
flare (left) and after the flare (right). The spectra are
very similar in overall shape, intensity, and resulting best
fit model parameters.  After the flare,
however, a significant excess of emission at 6.4 keV is
present which is not visible in the data before the flare (adapted from
\cite{ss_Gia+07b}).}
\label{figEl29_ss} 
\end{figure*}

Thanks to
{\em DROXO} it has been possible to perform, for the first
time, a detailed time-resolved study of the Fe 6.4 keV fluorescent line emission
of Elias 29 (\cite{ss_Gia+07b}).  The line intensity is highly
variable. It is absent at the beginning of observation, then after a quite
typical flare
(a factor 8 in intensity with a 6 ksec decay time) it appears
with a conspicuous 
equivalent width, EW $\sim$ 250 eV (cf. Fig. \ref{figEl29_ss}). Subsequently it continues to be present
with EW $\sim$ 150 eV
for the remaining 300 ksec (i.e. for 4 days!) of the observation. Apart for
the flare, the relatively soft X-ray spectra of Elias 29 remains essentially unchanged across
the entire observation, with no obvious hardening of the spectrum during 
the last 300 ksec of observation.
This behavior clearly challenges the "standard" interpretation of the
fluorescent emission being due to photo-ionizing X-ray photons (requiring
an adequate flux of photons with E $>$ 7.1 keV, that in this case seems to lack) and suggests an
alternative scenario in which the line is collisionally excited
by beams of electrons due to reconnections of magnetic field lines
occurring in the accretion funnels. The required energy can be released by magnetic fields stressed near the 
corotation radius as a result of the radial gradient of rotational speed.
%Such funnels, predicted from a long time by magnetospheric accretion model and, more recently, by MHD simulations of disk-star system, 
%are in agreement with the extent of some of flaring structures
%found in a dozen COUP and DROXO YSOs. 

Further long look continuous observations will permit us to 
verify if the phenomenon we have discovered in Elias~29 is present/common in other YSOs. Based on further {\em DROXO} results as well as on a recent 
time-resolved spectral analysis of the COUP data (Flaccomio 2007) I have not the space to discuss here, I am convinced that the YSO emission of the 6.4 keV  fluorescent line is a more complex phenomena that originally though and I am expecting soon further developments on this matter. On a somewhat longer time scale the next generation of imaging X-ray observatories (Simbol-X, XEUS, etc.) covering the bandpass up to $\sim$ 60 keV will allow us to directly probe the existence of a population of non-thermal electrons that is a key ingredient of the interpretational scenario proposed by Giardino and collaborators (2007b). 

\begin{figure}
\centering
\includegraphics[width=6cm]{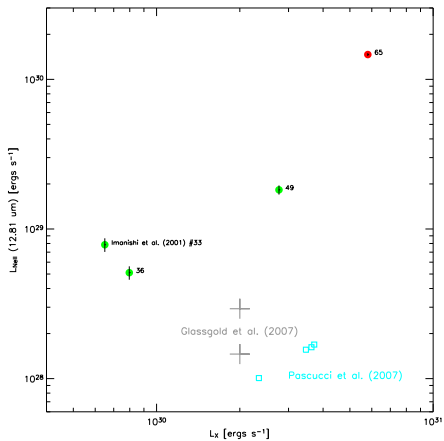}
\caption{Scatter plot of 12.81 ${\mu}$n IR line vs. X-ray luminosity of 
$\rho$ Oph YSOs. The data point from Pascucci et al. (2007) and the model prediction of Glassgold et al. (2007) are also shown.}
\label{figNeIR_ss} 
\end{figure}

Let me conclude by adding another piece of (controversial ?) evidence. The existence of an X-ray excited Ne 12.81 $\mu$m IR line has been predicted 
(Glassgold et al. 2007) and its detection has been ‏reported in 4 YSOs (Pascucci et al. 2007). ‏Using {\em Spitzer} archive spectra this line has been detected also in 4 $\rho$ Oph YSOs observed in X-rays, for 3 of which we have {\em DROXO} EPIC spectra (\cite{ss_Flacc+07b}). The X-ray brightest of them shows also the 6.4 keV Fe fluorescent line.‏ 
Fig. \ref{figNeIR_ss} shows a summary of the Pascucci et al. (2007) and of the new $\rho$ Oph data together with the model prediction. The $\rho$ Oph Ne IR luminosities are more than one dex higher than those of the somewhat older YSOs studied by Pascucci et al (2007). As of today we have no explanation to offer for this fact except to note the different age between the two groups of YSOs.

\section{Summary and Perspectives}
There is growing evidence of “interactions” between X-rays and YSO circumstellar disks:  big flares and the inferred long magnetic structures, the Fe 6.4 keV fluorescent line, the X-ray excited Ne  12.81 $\mu$m line, etc..  

Our understanding of the formation mechanism(s) of Fe 6.4 keV line is still limited and controversial. {\em DROXO} time-resolved spectroscopy of the Fe line challenges the “standard” Fe 6.4 keV formation scenario involving a direct interaction between the X-rays and disk material, but at the same time the detection of Ne (IR) line requires such an interaction. 

More long look observations and time-resolved spectroscopy are needed; 2-3 such XMM-Newton programs on nearby SFRs of, at least, 0.5 Msec each will serve this scope.

\begin{acknowledgements}
DROXO is a XMM-Newton Large Program (PI: S. Sciortino) 
supported by ASI-INAF contract I/023/05/0.
S. Sciortino acknowledges enlightening discussions
with many colleagues involved in the COUP and DROXO projects and their kind
share of results and figures in advance of final publication.
\end{acknowledgements}

\end{document}